\documentclass[aps,prl,superscriptaddress,twocolumn]{revtex4-1}

\usepackage{graphicx}
\usepackage[utf8]{inputenc}
\usepackage[english]{babel}
\usepackage{amsmath}
\usepackage{amsfonts}
\usepackage{amssymb}
\usepackage{amsbsy}
\usepackage{comment}
\usepackage{multirow}
\usepackage[yyyymmdd]{datetime}
\usepackage{booktabs}
\AtBeginDocument{
  \heavyrulewidth=.08em
  \lightrulewidth=.05em
  \cmidrulewidth=.03em
  \belowrulesep=.65ex
  \belowbottomsep=0pt
  \aboverulesep=.4ex
  \abovetopsep=0pt
  \cmidrulesep=\doublerulesep
  \cmidrulekern=.5em
  \defaultaddspace=.5em
}

\usepackage{setspace}

\usepackage{xcolor}
\definecolor{mygreen}{rgb}{0,0.6,0}
\definecolor{myblue}{rgb}{0.1,0,0.8}
\definecolor{myred}{rgb}{0.9,0.2,0.2}
\usepackage[colorlinks=true,bookmarks=true,pdfborder={0 0 0},
    linkcolor=myblue,urlcolor=myblue,citecolor=myblue,
    breaklinks=true,bookmarksnumbered=true]{hyperref}
\usepackage{colortbl}

\begin{document}

\author{D.~Santiago-Gonzalez}
\affiliation{Department of Physics and Astronomy, Louisiana State University, Baton Rouge, LA  70803, USA}
\affiliation{Physics Division, Argonne National Laboratory, Argonne, IL 60439, USA}

\author{K.~Auranen}
\affiliation{Physics Division, Argonne National Laboratory, Argonne, IL 60439, USA}

\author{M.~L.~Avila}
\affiliation{Physics Division, Argonne National Laboratory, Argonne, IL 60439, USA}

\author{A.~D.~Ayangeakaa}
\altaffiliation[Present address: ]{Department of Physics, United States Naval Academy, Annapolis, Maryland 21402, USA}
\affiliation{Physics Division, Argonne National Laboratory, Argonne, IL 60439, USA}

\author{B.~B.~Back}
\affiliation{Physics Division, Argonne National Laboratory, Argonne, IL 60439, USA}

\author{S.~Bottoni}
\altaffiliation[Present address: ]{Universit\`a degli Studi di Milano and INFN sez. Milano, I-20133, Milano, Italy}
\affiliation{Physics Division, Argonne National Laboratory, Argonne, IL 60439, USA}

\author{M.~P.~Carpenter}
\affiliation{Physics Division, Argonne National Laboratory, Argonne, IL 60439, USA}

\author{J.~Chen}
\affiliation{Physics Division, Argonne National Laboratory, Argonne, IL 60439, USA}

\author{C.~M.~Deibel}
\affiliation{Department of Physics and Astronomy, Louisiana State University, Baton Rouge, LA  70803, USA}

\author{A.~A.~Hood} 
\affiliation{Department of Physics and Astronomy, Louisiana State University, Baton Rouge, LA  70803, USA}

\author{C.~R.~Hoffman}
\affiliation{Physics Division, Argonne National Laboratory, Argonne, IL 60439, USA}

\author{R.~V.~F.~Janssens}
\altaffiliation[Present address: ]{Dept. of Physics and Astronomy, University of North Carolina at Chapel Hill, Chapel Hill, North Carolina 27599-3255, and Triangle Universities Nuclear Laboratory, Duke University, Durham, North Carolina 27708-2308, USA.}
\affiliation{Physics Division, Argonne National Laboratory, Argonne, IL 60439, USA}

\author{C.~L.~Jiang}
\affiliation{Physics Division, Argonne National Laboratory, Argonne, IL 60439, USA}

\author{B.~P.~Kay}
\affiliation{Physics Division, Argonne National Laboratory, Argonne, IL 60439, USA}

\author{S.~A.~Kuvin}
\affiliation{Department of Physics, University of Connecticut, Storrs, CT 06269, USA}

\author{A.~Lauer}
\affiliation{Department of Physics and Astronomy, Louisiana State University, Baton Rouge, LA  70803, USA}

\author{J.~P.~Schiffer}
\affiliation{Physics Division, Argonne National Laboratory, Argonne, IL 60439, USA}

\author{J.~Sethi}
\affiliation{Department of Chemistry and Biochemistry, University of Maryland, College Park, MD 20742, USA}
\affiliation{Physics Division, Argonne National Laboratory, Argonne, IL 60439, USA}

\author{R.~Talwar}
\affiliation{Physics Division, Argonne National Laboratory, Argonne, IL 60439, USA}

\author{I.~Wiedenh\"over}
\affiliation{Department of Physics, Florida State University, Tallahassee, FL 32306, USA}

\author{J.~Winkelbauer}
\affiliation{Los Alamos National Laboratory, Los Alamos, NM 87544, USA}

\author{S.~Zhu}
\affiliation{Physics Division, Argonne National Laboratory, Argonne, IL 60439, USA}

\title{Probing the single-particle character of rotational states in $^{19}$F using a short-lived isomeric beam}

\date{\today}

\begin{abstract}
A beam containing a substantial component of both the $J^{\pi}=5^+$,
$T_{1/2}=162$ ns isomeric state of $^{18}$F and its $1^+$, 109.77-min
ground state has been utilized to study members of the ground-state
rotational band in $^{19}$F through the neutron transfer reaction
$(d$,$p)$ in inverse kinematics.  The resulting spectroscopic
strengths confirm the single-particle nature of the 13/2$^+$
band-terminating state. The agreement between shell-model
calculations, using an interaction constructed within the $sd$ shell,
and our experimental results reinforces the idea of a
single-particle/collective duality in the descriptions of the
structure of atomic nuclei.
\end{abstract}

\maketitle

The duality of the collective and single-particle descriptions of the
structure of atomic nuclei has been recognized for some 60 years. It
is perhaps best summarized by an excerpt from the Nobel lecture of
Aage Bohr \cite{Lun92,Boh76}, \textit{``It was quite a dramatic moment
  when it was realized that some of the spectra in the light nuclei
  that had been successfully analyzed by the shell-model approach
  could be given a very simple interpretation in terms of the
  rotational coupling scheme.'' } Central to these comments by Bohr
was the ``\emph{special role}'' played by the $^{19}$F nucleus, one of
the lightest nuclei to exhibit rotational features. At that time, the
$^{19}$F spectrum had just been described by both shell-model
calculations assuming only a small number of valence
nucleons~\cite{Ell55,Red55}, as well as by a collective model assuming
rotational structures~\cite{Pau957,Rak57}.

Since these pioneering calculations, a great deal of work has been
done aiming to refine the theoretical description of
$^{19}$F~\cite{Red58,Ari68,Ham67,Ben69,Oya75,Bro85,Bro1988,Til95,USDAB},
not the least of which involved identifying the similarities between
wave functions generated from both approaches~\cite{Ell58} and the
recent accessibility of $^{19}$F to $\emph{ab initio}$
calculations~\cite{Str16}. Within a collective description, the
ground-state rotational band in $^{19}$F exhibits the characteristic
staggering, or ``signature splitting,'' of a $K=1/2$ rotational
structure~\cite{BM2} where a measured static quadrupole moment points
towards prolate deformation and the states are linked by relatively
enhanced electric quadrupole transitions.  The band proceeds from its
bandhead, with a spin-parity of $J^{\pi}_{\text{min}}$=1/2$^+$, to its
terminating state, $J^{\pi}_{\text{max}}=13/2^{+}$. This termination
is evidence for the importance of shell structure since this is the
maximum spin that can be generated from three nucleons in the
$sd$-shell outside the $^{16}$O core.

The nucleus $^{18}$F has a $J^{\pi}=5^+$ excited state that consists
of two maximally-aligned $0d_{5/2}$ nucleons outside the closed-shell
$^{16}$O core~\cite{Pol69}. This level has a $162(7)$-ns
half-life~\cite{Til95}, comparable to the flight time of an ion beam
in the tens of MeV/u range over a few meters. By producing a beam of
this isomer ($^{18m}$F), and bringing it to a target, states in
$^{19}$F of $J\ge 5/2$, including the 13/2$^+$ terminating state,
can be produced by the addition of yet another $0d_{5/2}$ neutron.

In this Letter, we report on the extraction of spectroscopic overlaps
between initial states in $^{18m}$F and final states in $^{19}$F,
including the terminating 13/2$^+$ state of the $K=1/2$ rotational
band, via the single-neutron ($d$,$p$) transfer reaction. Combined
with a simultaneous measurement of the ($d$,$p$) reaction with a
$^{18}$F beam in its $J^{\pi}=1^+$ ground state ($^{18g}$F), whereby
the lower-spin members of the band were populated, a determination of
the single-particle character of the $^{19}$F ground-state rotational
band was obtained for the first time in a single experiment.

The experiment was performed at the ATLAS facility at Argonne National
Laboratory and utilized the HELIOS spectrometer~\cite{Wuo07,Lig10}, a
device designed for measuring transfer reactions in inverse
kinematics. A radioactive beam of $^{18}$F at an energy of 14~MeV/u
was produced with an intensity of $\sim5\times10^5$~pps via the
in-flight technique~\cite{Har00,Reh11}. The
$^{2}$H($^{17}$O,$^{18}$F)$n$ production reaction was used with an
$^{17}$O primary beam (15~MeV/u) at a typical intensity of 60~pnA. A
cryogenically-cooled deuterium-filled gas cell ($\sim$80~K and
1.4$\times$10$^5$~Pa) provided the production target material. The
resulting $^{18}$F beam was comprised of ions in both ground and
isomeric states. Previous experiments using $^{18m}$F beams include
those of Refs.~\cite{Bec90,Bro95,Rob95,Rob02,Zim07}. In the present
work, the $^{18m}$F/$^{18g}$F ratio has been estimated to be 0.56(8)
immediately after production and 0.11(2) after transport to the HELIOS
experimental station (details on this estimation are given below).

\begin{figure}
\centering
\includegraphics[width=0.54\textwidth]{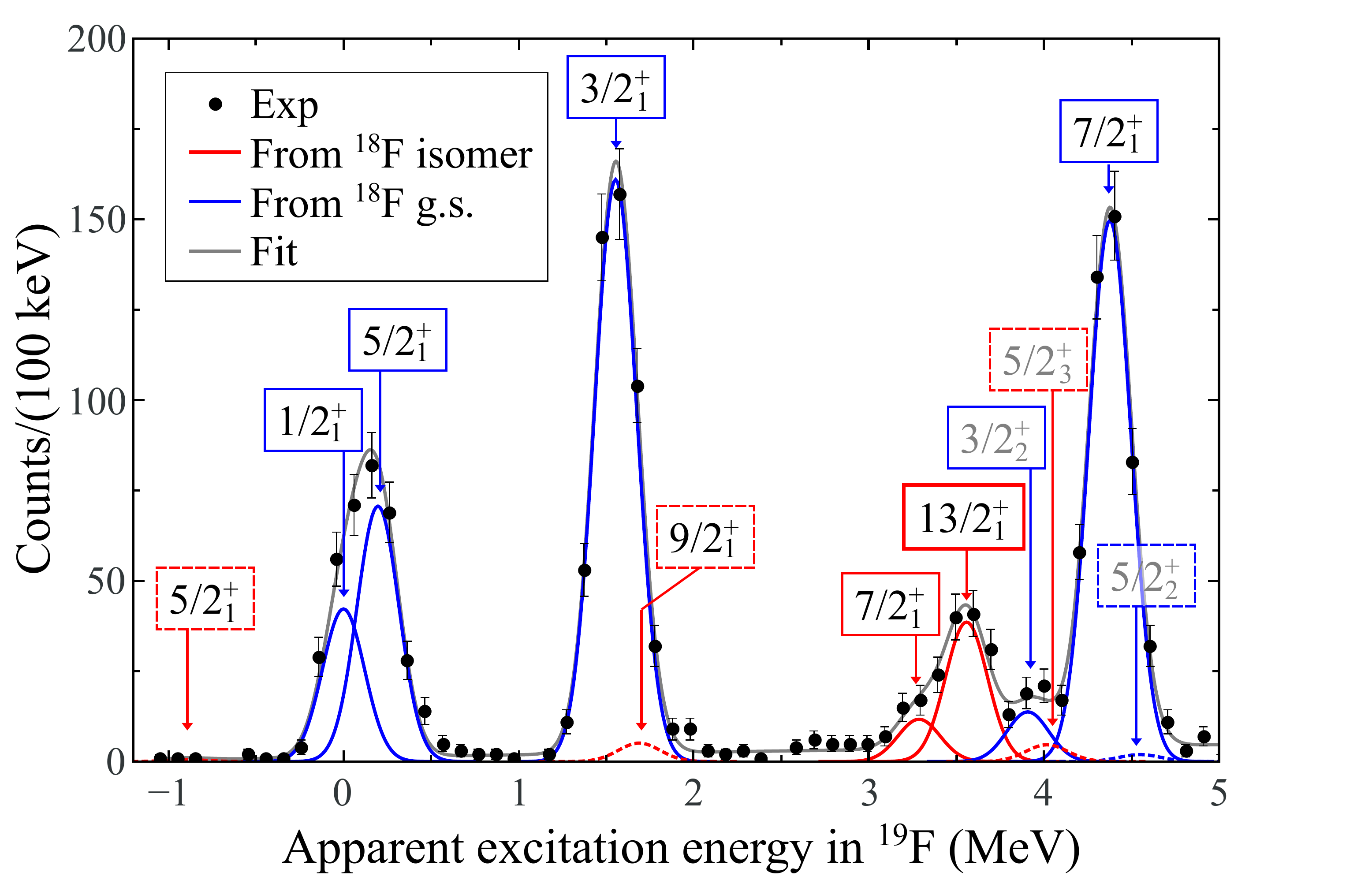}
\caption{
  Apparent excitation energy in $^{19}$F extracted from protons in
  coincidence with $^{19}$F recoils following $^{18g,m}$F($d$,$p$)
  reactions (black points with statistical uncertainties). A
  multi-Gaussian fit of the known levels in $^{19}$F including a small
  linear background and fixed widths is shown in gray. States
  populated from ($d$,$p$) reactions on $^{18g}$F are in blue while
  those from $^{18m}$F are in red. Weak levels, which, if removed from
  the fit would have little effect on the $\chi^2$ value are
  represented by dashed lines. Black labels identify states belonging
  to the ground-state rotational band.}
\label{F:ex}
\end{figure}

HELIOS was configured for the observation of protons in coincidence
with $^{19}$F from single-neutron transfer reactions, ($d$,$p$), on
beams of both $^{18g}$F and $^{18m}$F.  The solenoidal field was set
to 2.85~T and deuterated polyethylene (CD$_2$) targets with a nominal
thickness of 400~$\mu$g/cm$^2$ were placed near the center of the
field region. Upstream of the target location, an on-axis
position-sensitive Si detector array was installed for proton
detection. Protons were uniquely identified from their cyclotron
periods after completing a single orbit from the target to the Si
detector array. A fast-counting, segmented ionization
chamber~\cite{Lai17} centered around 0$^{\circ}$ was positioned
downstream of the target for $^{19}$F recoil detection. Coincidence
events between protons and recoiling ions were determined by the
relative time difference between the two detectors. Acceptance for
proton-recoil events was possible up to $\sim$5~MeV in excitation
energy, covering all but the $11/2^+_1$ member in the $^{19}$F
ground-state rotational band. The acceptance also included proton
center-of-mass angles, $\theta_{\text{c.m.}}$, ranging from
$\sim$10-35$^{\circ}$.

Levels in $^{19}$F populated by reactions on the isomeric beam appear
shifted by -1.07~MeV relative to ground-state reactions, hence the
`apparent' qualifier in the angle-integrated excitation spectrum of
Fig.~\ref{F:ex}. The shift is primarily the result of the $Q$-value
difference between $^{18m}$F($d$,$p$) ($Q$ = 9.328~MeV) and
$^{18g}$F($d$,$p$) ($Q$ = 8.207~MeV). In addition, a $\sim$50~keV
shift arises from differences in the kinematics between the two
reactions. The $Q$-value resolution was 280~keV FWHM, driven primarily
by the target thickness and the emittance of the secondary beam. The
best fit to the data using known $^{19}$F excitation
energies~\cite{Til95} is shown in Fig.~\ref{F:ex} by the solid grey
line.

\begin{figure}
\centering
\includegraphics[width=0.45\textwidth]{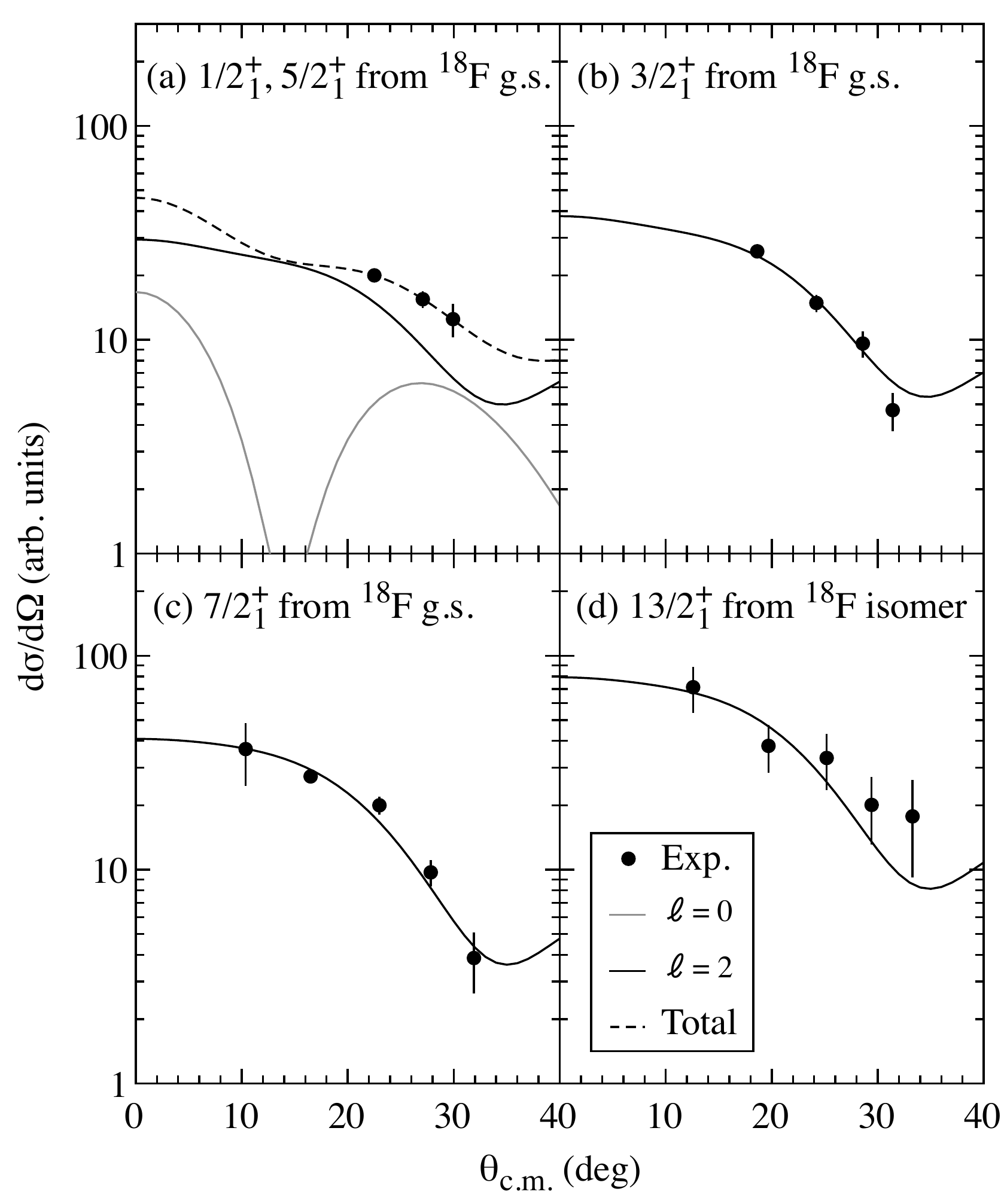} 
\caption{
  Angular distributions for states in $^{19}$F obtained from
  $^{18g}$F$(d,p)$ reactions, (a) $1/2^+_1$ and $5/2^+_1$ doublet, (b)
  $3/2^+_1$, (c) $7/2^+_1$, and from $^{18m}$F$(d,p)$ reactions, (d)
  $13/2^+_1$. The $13/2^+_1$ data include the 0.11(2) normalization
  factor to account for the $^{18m}$F/$^{18g}$F secondary beam
  ratio. The DWBA calculations are represented by the lines.}
\label{F:ad}
\end{figure}

Angle-integrated cross sections were determined from measured yields
for all states identified in Fig.~\ref{F:ex}. For the levels that were
populated strongly, relative differential cross sections,
$d\sigma/d\Omega$, and angular distributions were also derived and are
presented in Fig.~\ref{F:ad}. The center-of-mass angle,
$\theta_{\text{c.m}}$, for each data point in Fig.~\ref{F:ad}
corresponds to the average angle covered by one set of
position-sensitive Si detectors and has an uncertainty of
$\lesssim0.5^{\circ}$. Upper limits on yields were determined for
weaker states by an increase of 5\% to the best-fit $\chi^2$ value to
the apparent excitation spectrum (Fig.~\ref{F:ex}). The cross section
for levels populated by the isomeric component of the beam were
corrected for the $^{18m}$F/$^{18g}$F beam ratio at the HELIOS target.

The relative intensity of the isomeric 5$^+$ component of the $^{18}$F
beam compared to the 1$^+$ ground state was estimated by calculating
the populations at the gas cell and then accounting for the reduction
in the ratio over the flight time of the beam to the target at the
HELIOS experimental station. There are no experimental data available
for the ($d$,$n$) beam-production reaction at the relevant energies
and, therefore, the relative strengths of the population of the bound
states of $^{18}$F were taken from an analogous $^{17}$O($^3$He,$d$)
proton-transfer reaction~\cite{Pol69}. The bulk of the relevant
reaction yield proceeds to ten states below the proton separation
energy in $^{18}$F. The high-lying states decay by prompt $\gamma$-ray
emission with known branching ratios to either $^{18g}$F or
$^{18m}$F~\cite{Til95}. The $^{2}$H($^{17}$O,$^{18}$F)$n$ cross
sections were calculated with the distorted wave Born approximation
(DWBA) utilizing the {\sc Ptolemy} code~\cite{Pto78}. The DWBA
prescription, including the choice of optical-model parameters, was
validated through comparisons with available $^{16}$O($d$,$p$)
cross-section data at a similar energy (13.15~MeV/u)~\cite{Tes1964}.

The $^{18m}$F/$^{18g}$F ratio at the production gas cell was
calculated to be 0.56(8). The flight path from the gas cell to the
HELIOS experimental station was 16.3~m, and at a beam energy of
14~MeV/u, corresponds to a time of flight of $1.9 \,T_{1/2}$ of the
isomeric state. Hence, the ratio at the HELIOS target was
$^{18m}$F/$^{18g}$F~$=$~0.11(2). 

The relative single-neutron overlaps (spectroscopic factors) between
initial states in $^{18}$F and final states in $^{19}$F, $S$ (isospin
factor $C^2=1$), were extracted from the ratio of measured cross
sections to those calculated with DWBA. In the standard procedure, the
depth of the Woods-Saxon potential was varied to reproduce the binding
energy of each final state.  The deuteron wave function was calculated
with the V$_{18}$ potential~\cite{V18}. A global set of optical model
parameters~\cite{An06} was used to calculate the angular distributions
shown in Fig.\ref{F:ad}. Angular distributions obtained using a static
set of parameters~\cite{Sch67} produced similar results within
uncertainties.

\begin{figure}
\centering
\includegraphics[width=0.5\textwidth]{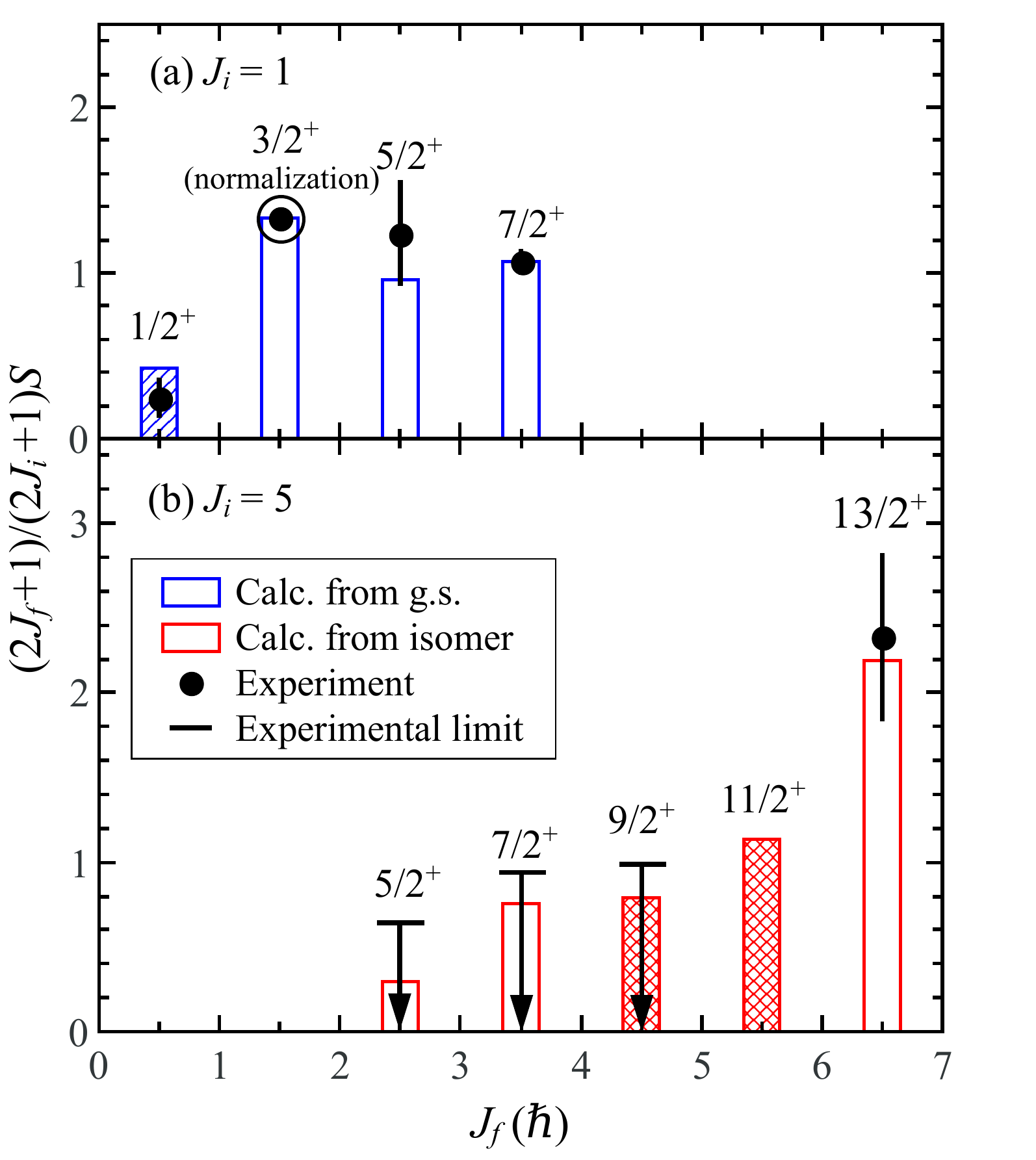} 
\caption{The information on relative strengths of states in $^{19}$F
  is plotted as a function of their spin, separately for the
  $^{18g}$F($d$,$p$) (a) and $^{18m}$F($d$,$p$) reaction (b). The
  limit on the $9/2^+$ state is obtained assuming
  $\ell=2$. Shell-model calculations using the USDB interaction are
  represented by bars for $\ell=0$ (striped), $2$ (open), or 0 \& 2
  (hatched) strengths. }
\label{F:rvj} 
\end{figure}

The $S$ values resulting from best fits of the DWBA calculations to
the angular distributions (Fig.~\ref{F:ad}), as well as upper limits
on the $S$ values determined from ratios of integrated cross sections,
are given in Table~\ref{T:R} while the spectroscopic strengths,
$(2J_i+1)/(2J_f+1)S$, are shown in Fig.~\ref{F:rvj}. All $S$ values
have been normalized to the $3/2^+_1$-$^{18g}$F transfer spectroscopic
factor. Uncertainties on $S$ are due to the choice of optical-model
and bound-state parameters of the DWBA calculations. Uncertainty in
the deduced $^{18m}$F/$^{18g}$F ratio of the beam also contributes for
levels which were populated by transfer on the isomer.

\begin{table}

  \caption{ \label{T:R} Relative spectroscopic factors, $S$, for
    levels belonging to the $^{19}$F $K=1/2^+$ band~\cite{Til95}. All
    $S$ values are normalized to that of the 1.554-MeV 3/2$^+_1$
    level. Only $S$ values above 0.01 are shown and (---) signifies the
    non-observation or inaccessibility of a given level.}

  \newcommand\T{\rule{0pt}{3ex}} 
  \newcommand \B{\rule[-1.5ex]{0pt}{0pt}}
  \begin{ruledtabular} 
    \begin{tabular}{ccccccc}
      & & & & \multicolumn{3}{c}{$S$} \\
      \cline{5-7} 
      \T\B $E_f$ (MeV) & $J^\pi_f$ & $J^\pi_i$ & $\ell$ & Present & Ref.~\cite{Koz06} & Theory\footnotemark[1] \\
      \hline  
      \T\B 0 & $1/2^+_1$ & $1^+$ & 0 & 0.4(2) & 0.75(15) & 0.64 \\
      \arrayrulecolor{lightgray}\cline{2-7} 
      \T \multirow{2}{*}{0.197} & \multirow{2}{*}{$5/2^+_1$} & $1^+$ & 2 & 0.6(2) & 0.40(8) &  0.48 \\
      \B & & $5^+$ & 2 & $<$1.0\footnotemark[2] & ---  &  0.54 \\
      \cline{2-7} 
      \T\B 1.554 & $3/2^+_1$ & $1^+$ & 2 & 1 & 1 & 1 \\
      \cline{2-7} 
      \T \multirow{2}{*}{2.780} & \multirow{2}{*}{$9/2^+_1$} & \multirow{2}{*}{$5^+$} & 0 & $<$0.4\footnotemark[2]\footnotemark[3] & --- & 0.30 \\
      \B & & & 2 & $<$1.2\footnotemark[2]\footnotemark[3] & --- & 0.57 \\
      \cline{2-7} 
      \T \multirow{2}{*}{4.378} & \multirow{2}{*}{$7/2^+_1$} & $1^+$ & 2 & 0.40(3) & 0.5(1) & 0.39 \\
      \B & & $5^+$ & 2 & $<1.3$\footnotemark[2] & ---   & 1.03 \\
      \cline{2-7} 
      \T\B 4.648 & $13/2^+_1$&$5^+$ & 2 & 1.8(4)\footnotemark[2] & --- & 1.72 \\
      \cline{2-7} 
      \T \multirow{2}{*}{6.500} & \multirow{2}{*}{$11/2^+_1$} & \multirow{2}{*}{$5^+$} & 0 & --- & --- & 0.50 \\
      \B & & & 2 & --- & --- & 0.54 \\
      \arrayrulecolor{black}
    \end{tabular}
  \end{ruledtabular}
  \footnotetext[1]{Shell-model calculations using the USDB interaction~\cite{USDAB}.}
  \footnotetext[2]{Includes calculated value of 0.11(2) for the
    $^{18}$F isomer to g.s. ratio in the secondary beam.}
  \footnotetext[3]{Assumed pure $\ell$ transfer.}
\end{table}

The data on the population of the $^{19}$F $K=1/2^+$ band are clearly
present in the apparent excitation energy spectrum of Fig.~\ref{F:ex}.
As expected from the relatively small isomeric component in our beam,
the dominant features in our spectrum are similar to those in Fig.~2
of Ref.~\cite{Koz06} and Fig.~5 of Ref.~\cite{Ser07}, where there was
no isomeric component in the beam. The $S$ values deduced for the
lower-spin members of the $K=1/2^+$ band populated by transfer on the
$^{18g}$F, namely the $1/2^+$ (0.000~MeV), $5/2^+$ (0.197~MeV),
$3/2^+$ (1.554~MeV), and $7/2^+$ (4.378~MeV) $^{19}$F states, are
consistent with those from Ref.~\cite{Koz06} (see Table~\ref{T:R}).

For the unresolved lowest-lying $1/2^+$ and $5/2^+$ levels, single line
shapes with $\ell=0$ and $2$ transfers were assumed, respectively. Due
to the limited angular coverage, our measurement is not sensitive to
the population of the 0.110-MeV, 1/2$^-_1$ level. The angular
distributions of the $3/2^+$ and $7/2^+$ states did not require any
sizable contributions ($>5$\%) from $\ell=0$ neutron transfer.

There are structures in the spectrum of Fig.~\ref{F:ex}, noticeable
between 3-3.8~MeV, a featureless region in the $^{18g}$F transfer
spectra of Refs~\cite{Koz06,Ser07}. Accounting for the -1.07-MeV shift
in apparent excitation energy for the ($d$,$p$) reaction on $^{18m}$F,
there are three previously known levels in this region that are
accessible via $\ell=0$ or 2 neutron transfer: the $7/2^+_1$
(4.378~MeV), $5/2^+_2$ (4.550~MeV), and $13/2^+_1$ (4.648~MeV)
states~\cite{Til95}. Indeed, in the apparent energy spectrum of
Fig.~\ref{F:ex}, lines corresponding to the population of the $13/2^+$
and $7/2^+$ levels are observed in the 3-3.8~MeV range, identifying
neutron transfer onto the isomeric 5$^+$ level of $^{18}$F for the
first time. Of the five other known levels also open to population
through transfer on $^{18m}$F in the energy region covered, upper
limits on yields for the $5/2^+_1$ (-0.873~MeV), $9/2^+_1$
(1.710~MeV), and $5/2^+_3$ (4.037~MeV) states could be determined. The
angular distribution for the $13/2^+$ state, and the resulting DWBA
fit [Fig.~\ref{F:ad}(d)], identify it as a strong $\ell=2$ neutron
transfer, solidifying its population from $^{18}$F in its 5$^+$
isomeric state.

Accessibility to an in-flight beam of $^{18}$F in both its ground
$1^+$ and fully stretched 5$^+$ states has enabled the extraction of
(or setting limits on) the relative spectroscopic overlaps of the
$1/2^+,3/2^+,5/2^+,7/2^+,9/2^+$ and 13/2$^+$ members of the
ground-state rotational band of $^{19}$F (Table~\ref{T:R} and
Fig.~\ref{F:rvj}). The extracted $S$ value for the $13/2^+$ state, and
its spectroscopic strength exceed those of all other states in the
rotational band. This observation confirms the dominant
single-particle configuration in this band-terminating state as
corresponding to the maximally-aligned state with a
$\pi(0d_{5/2})^1_{J=5/2}\otimes\nu(0d_{5/2})^2_{J=4}$
configuration. This is the first direct measurement of the
single-particle nature of a high-spin terminating state. This result,
together with the strengths of the levels populated from $^{18g}$F,
and the upper limits on the strengths of states populated from
$^{18m}$F, describe the evolution of the single-particle strength of
the states in the rotational band as a function of spin, from
inception to termination (see Fig~\ref{F:rvj}).

Comparisons between the extracted $S$ values and strengths of the
present work to those calculated by the $sd$-confined USDB
interaction~\cite{USDAB} are also given in Table~\ref{T:R} and
Fig.~\ref{F:rvj}. The calculations are consistent with the
experimental values, or limits, even though these incorporate only
three valence particles (one proton and two neutrons) and three active
orbitals for each nucleon.

The present results highlight the single-particle character of the
highest-spin state ($13/2^+$) in the rotational band of $^{19}$F by
confirming that the associated configuration corresponds to the
maximally-aligned, terminating state. Furthermore, we have found that
the spectroscopic factors from shell-model calculations are consistent
with our experimental values (and limits) for the $^{19}$F states
members of the ground-state rotational band. Hence, some 40 years
after his seminal statements~\cite{Lun92,Boh76}, A. Bohr's dual
interpretation of the $^{19}$F sequence in terms of a collective
and/or a single-particle excitation has been reinforced.


In summary, the single-particle character of members belonging to the
ground-state rotational band in $^{19}$F, including the terminating
$13/2^+$ state, have been probed in a single measurement via the
($d$,$p$) reaction. The relatively large spectroscopic strength
observed for the $13/2^+$ level confirms the wave function purity
expected in a maximally-aligned, terminating state. Agreement between
shell-model calculation and the experimentally determined
spectroscopic factors for the inspected rotational states strengthens
the notion of a collective and single-particle duality in the
descriptions of the structure of atomic nuclei.  The present
measurement was possible only through the production of a beam of
$^{18}$F whereby a significant fraction of ions resided in their
short-lived isomeric state.

\section{Acknowledgments}
The authors thank C.~Dickerson and R.~C.~Pardo for developing the
$^{18}$F isomeric beam.  This material is based upon work supported by
the U.S. Department of Energy, Office of Science, Office of Nuclear
Physics, under contract number DE-AC02-06CH11357 (ANL) and grant
No. DE-FG02-96ER40978 (LSU).  This research used resources of ANL's
ATLAS facility, which is a DOE Office of Science User Facility.
\singlespacing

\bibliographystyle{apsrev4-1}
\bibliography{References}

\end{document}